\documentclass{pasj00}

\SetRunningHead{S. Konami et al.}{
Abundance patterns in the ISM of NGC~3079}

\title{
Spatial Distribution of Abundance Patterns \\
in the Starburst Galaxy NGC~3079 Revealed with Chandra and Suzaku}

\author{
 Saori \textsc{Konami},\altaffilmark{1,2}
 Kyoko \textsc{Matsushita},\altaffilmark{1} 
 Poshak \textsc{Gandhi},\altaffilmark{3} 
and Toru \textsc{Tamagawa},\altaffilmark{2,1}
}

\altaffiltext{1}{
Department of Physics, Tokyo University of Science,
1-3 Kagurazaka, Shinjuku-ku, Tokyo 162-8601}
\altaffiltext{2}{
High Energy Astrophysics Laboratory, the Institute of Physical and Chemical Research,
2-1 Hirosawa, Wako, Saitama 351-0198}
\email{konami@crab.riken.jp}
\altaffiltext{3}{
Institute of Space and Astronautical Science (ISAS), Japan Aerospace Exploration Agency, 
3-1-1 Yoshinodai, Chuo-ku, Sagamihara, Kanagawa 252-5210}

\Received{}
\Accepted{}
\Published{---}

\begin{document}
\maketitle
\begin{abstract}

We performed simultaneous spectral analysis of 26.6 ksec of Chandra and 102.3 ksec of 
Suzaku X-ray data of the starburst galaxy NGC~3079. 
The spectra are extracted from four regions: 
0.5$'$ (2.25~kpc) circle, an inner 0.5$'$-1$'$ (2.25--4.5~kpc) ring, and an outer 1$'$-2$'$ (4.5--9~kpc) 
ring from Chandra, and 4$'$ (18~kpc) circle from Suzaku, all centered on the nucleus. 
Fitting with thermal plasma models yields interstellar medium (ISM) temperatures of 
0.65$^{+0.05}_{-0.04}$, 0.45$^{+0.07}_{-0.06}$, and 0.24$^{+0.03}_{-0.02}$~keV in the three regions, respectively. 
The combination of Chandra's high angular resolution and Suzaku's good spectral sensitivity 
enable us to spatially resolve and measure the abundances of the metals 
O, Ne, Mg, and Fe within the hot ISM.
In particular, the abundance patterns of O/Fe, Ne/Fe, Mg/Fe, and Si/Fe in the central regions ($<$4.5~kpc) are 
consistent with the expectations from Supernovae (SN) II synthesis. 
On the other hand, the pattern in the region beyond 4.5~kpc is closer to solar. 
The central regions are also where copious polycyclic aromatic hydrocarbon infrared 
emission related to recent starburst activity is known to occur. 
This suggests that we are seeing starburst-related SN II metal enrichment in the hot 
X-ray--emitting nuclear ISM. 
The spatial extent of the SN II--like abundance patterns is consistent with NGC~3079 being 
in a relatively-early phase of starburst activity.

\end{abstract}

\section{Introduction}

\begin{figure*}
\begin{center}
\centerline{
\FigureFile(\textwidth,\textwidth){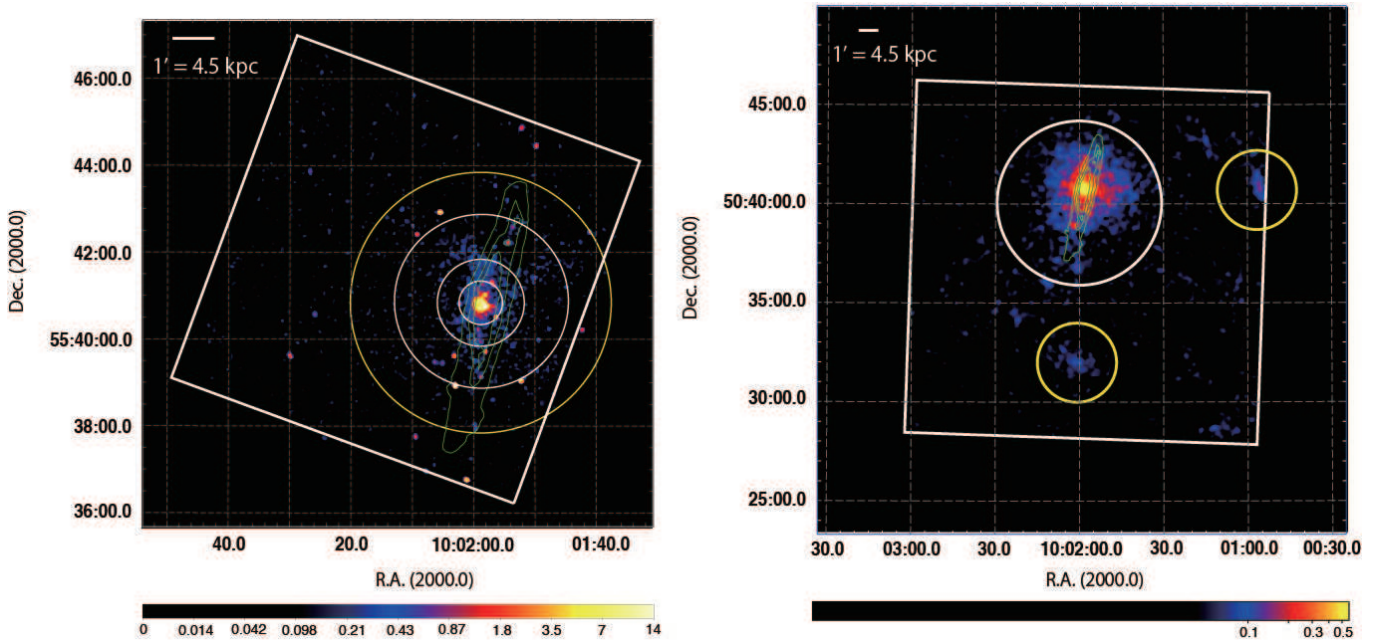}}
\caption{
The left and right-hand panels are the X-ray images of NGC~3079 with Chandra ACIS-S3 and with Suzaku XIS1, 
respectively. 
Background components were not subtracted and vignetting was not corrected for in these images. 
These images are overlaid with optical galactic disk contours from the Digitized Sky Survey (6450\AA) in green. 
Left; the ACIS-S3 image in 0.3--5.0~keV is smoothed with a $\sigma$=1.95$''$ Gaussian profile. 
The radii of the white circles are 0.5$'$, 1$'$, 2$'$, and 3$'$ moving out from the nucleus. 
The white box shows the ACIS-S3 chip boundary. 
Right; the XIS~1 image in 0.5--5.0~keV is smoothed with a $\sigma$=6.24$''$ Gaussian profile. 
The radius of the white circular source region is 4$'$. 
Background spectra were extracted from the full Suzaku XIS field of view excluding the source aperture 
and the two yellow circles centered on field sources. 
The white box is the Suzaku XIS field of view. 
}\label{fig_ima}
\end{center}
\end{figure*}

The hot interstellar medium (ISM) in galaxies hold important information to 
reveal the chemical evolution of galaxies and clusters of galaxies. 
O and Mg are synthesized by type II supernovae (SN II), while Fe is mainly produced 
by type Ia supernovae (SN Ia). 
Therefore, the ratios of metal abundances provide us clues to investigate 
the star formation history in galaxies. 
In addition, starburst galaxies play a prominent role in the chemical 
evolution of the universe since outflows inject metals into the intra-cluster 
medium (ICM) and warm-hot intergalactic medium (WHIM). 

Observations of starburst galaxies in X-rays have recently been performed to measure 
the metal abundance patterns in the ISM. 
XMM-Newton detected the metals O, Ne, Mg, Si, and Fe at $\lesssim$ 3~kpc from the nucleus 
along the outflow axis of NGC~253 \citep{bauer_07}. 
For M~82, the abundance patterns were derived in the outflow to `cap' regions located 
$\sim$11 kpc to the North (\cite{tsuru_07}; \cite{konami_11}). 
The ratios of O/Fe, Ne/Fe, and Mg/Fe are higher than the solar values 
by factors of 2--3, as expected from SN II yields. 
Furthermore, for NGC~4631, the outflow wind and disk regions show patterns close to 
those expected from the SN II yields and solar abundance \citep{lodders_03}, respectively \citep{yamasaki_09}. 
These results support the scenario where starburst activity enriches the outflow through SN II metal ejection 
into intergalactic space. 
In contrast, NGC~4258 is a non-starburst spiral galaxy, whose ISM abundance pattern 
is fully consistent with solar \citep{konami_09}. 
In fact, spiral galaxies at the present cosmic epoch commonly show solar abundance ratios 
unaffected by starburst activity.

NGC~3079 is a nearby SB(s)c starburst galaxy with a high inclination angle of 84$^\circ$ 
and a disk position angle of 166$^\circ$ \citep{irwin_91} determined in the optical. 
Therefore, this galaxy gives a clear view along the outflow axis perpendicular to the disk, 
suitable for investigating chemical enrichment by the starburst. 
The derived radial velocity of the CO emission line imply the existence of 
an ultra-high-density core of molecular gas 
within a radius of 125~pc, with mass as large as $\sim3\times10^8~M_{\odot}$ \citep{sofue_01}. 
In the far-infrared, the luminous and extended emission is distributed across the entire disk of the galaxy 
rather than being concentrated in an central region \citep{perez_00}. 
The observations of polycyclic aromatic hydrocarbons suggest the occurrence of 
active starformation in the central 4~kpc region \citep{yamagishi_10}.
The outflow with kiloparsec-scale along the minor axis have been observed in the radio 
(\cite{duric_88}; \cite{irwin_03}), H$\alpha$ (\cite{ford_86}; \cite{veilleux_94}), 
and X-ray emission (\cite{dahlem_98}; \cite{pietsch_98}). 
In particular, there is a tight correlation between H$\alpha$ and X-ray emission in the superbubble, 
but the nature of the power source of the superbubble is still debated, 
with both active galactic nucleus and starburst sources discussed \citep{cecil_02}. 
\citet{pietsch_98} reproduced the spectra of ROSAT X-ray data in the energy range of 0.1--2.4~keV 
by using thermal plasma models with a negative radial temperature gradient moving outwards.

NGC~3079 is also classified as a low-ionization nuclear emission-line region or as a Seyfert 2 galaxy 
in the optical. \citet{iyomoto_01} detected a heavily absorbed continuum 
and a strong Fe-K emission line, using BeppoSAX data in the energy range 2--100~keV. 
They estimated that the contributions of the AGN and the starburst to the energy output of 
NGC~3079 are of the same order of magnitude. 
\citet{fukazawa_11} systematically analyzed the Fe-K line features of Seyfert galaxies 
including NGC~3079'¡¡with Suzaku data. They revealed neutral matter causing a 6.4~keV line 
and 7.1~keV edge is attributed to the torus, constraining the ionization state and the relation 
among lines, column density, and edge depth.

In this paper, we investigate the spatial abundance patterns in the hot ISM of NGC~3079. 
The paper is structured as follows. 
In section 2, we summarize observations by Chandra and Suzaku. 
Chandra's sharp angular resolution allows spatial resolution of emission from the ISM and 
individual contaminating point sources. Suzaku's excellent spectral sensitivity allows 
high quality modeling of the abundance patterns. 
Section 3 details the data analysis and results, and section 4 gives a discussion of these results. 
Finally, in section 5, we present our conclusions.
We adopted 15.6~Mpc for the distance to NGC~3079 \citep{sofue_99}, where 1$'$ corresponds to 4.5 kpc.
Unless noted otherwise, we use the solar abundances from \citet{lodders_03}, 
and the quoted errors are for a 90\% confidence interval 
for a single interesting parameter.

\section{Observation and Data Reduction}
\label{sec:obs}

\subsection{Chandra}
\label{sub:chandra}

Chandra observed NGC~3079 (ObsID 2038) in 2001 March, using the Advanced CCD Imaging Spectrometer 
(ACIS; \cite{garmire_03}). 
The aimpoint on the default location of the S3 chip is (R.A., Dec.)=(\timeform{10h01m53.69s}, 
\timeform{+55D40'41.1''})\@.
We reprocessed all the level-1 event data, applying the standard data reduction using 
CIAO version 4.3 and CALDB version 4.3.3. 
The light curve over 0.4--10~keV are extracted with CIAO tool ``dmextract''. 
There were no anomalous events at greater or less than $3\sigma$ around the mean count rate, 
so the full exposure time of 26.6~ksec was retained.

\begin{table}
\caption{
Best-fit parameters for the integrated spectrum of point sources (including the nucleus) of Chandra with  apec components + PL model.$^{\ast}$}
\label{tab_point}
\begin{center}
\begin{tabular}{lccc} \hline\hline
Parameters               &             &                          \\ \hline
$N_{\rm H_{PS}}$               & (10$^{20}$ cm$^{-2}$) & 115$_{-35.4}^{+16.7}$    \\
$\Gamma_{\rm PS}$        &             & 1.92$_{-0.28}^{+0.15}$            \\ 
line energy            & (keV) & 6.30$_{-0.03}^{+0.41}$   \\ [2.0ex] 
${\chi^2}$/d.o.f.         &             & 46/36                   \\ \hline \hline
\end{tabular}
\end{center}
\end{table}

\begin{figure}
\begin{center}
\centerline{
\FigureFile(0.5\textwidth,0.5\textwidth){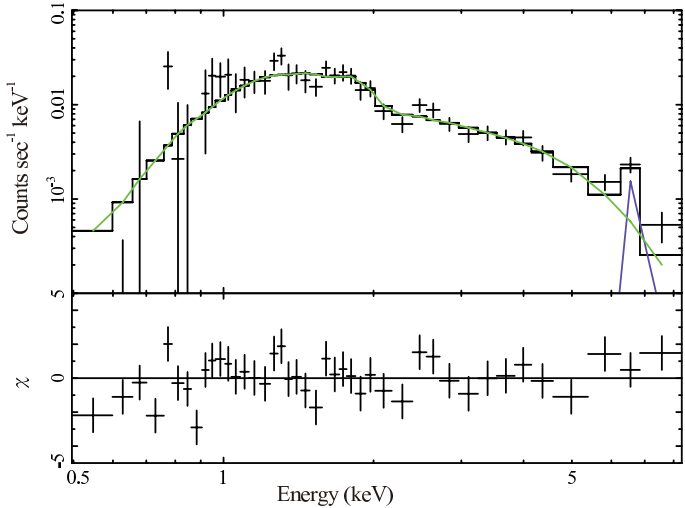}}
\caption{
Background-subtracted Chandra ACIS-S3 spectrum (black data) of accumulated emission 
for point sources, shown without removal of instrumental response. 
The black histogram shows the best-fit model from table \ref{tab_point}. 
The green line shows the absorbed power-law (PL$_{\rm PS}$) component, and 
the purple line is the Fe-line component (Gaussian).
}\label{fig_point}
\end{center}
\end{figure}

\subsection{Suzaku}
\label{sub:suzaku}
Suzaku performed an observation of NGC~3079 in 2008 May, with the target centered 
on the HXD aimpoint (ObsID 803039020). 
The averaged pointing direction is centered on (R.A., Dec.)=(\timeform{10h01m59.33s}, 
\timeform{+55D36'41.8''})\@.
The XIS consists of two front-illuminated CCD cameras (FI: XIS0 and XIS3), and 
one which is back-illuminated (BI: XIS1) \citep{koyama_07}. 
The XIS was operated in the normal 
clocking mode (8~s exposure per frame), with the standard 
5 $\times$ 5 and 3 $\times$ 3 editing mode. 
We processed the XIS data using the ``xispi'' and ``makepi'' ftools task 
and the CALDB files of version 2009-08-13. 
The XIS data were then cleaned by assuming thresholds on the Earth elevation angle 
of $>~5^{\circ}$ and the Day-earth elevation angle of $>~20^{\circ}$. 
We also discarded data for time since passage through the South Atlantic 
Anomaly of less than 436 sec. 
We created a light curve for each sensor over 0.5--2 keV binned to 540 sec 
but there are no anomalous event rate greater or less 
than $3\sigma$ around the mean. 
After the above screening, the remaining exposure was 102.3 ksec 
for both FIs and BI. 
Event screening on cut-off rigidity was not performed.

Spectral analysis was performed with HEAsoft version 6.10.
In order to subtract the non-X-ray background (NXB), we employed 
the dark-Earth database available through the ``xisnxbgen'' ftools task \citep{tawa_08}.
We generated two different Ancillary Response Files (ARFs) 
for the spectrum of each region, which assumed the source image as uniform 
sky emission and the observed XIS1 image by the ``xissimarfgen'' ftools 
task \citep{ishisaki_07}. We also included the effect of contamination
on the optical blocking filters of the XIS in the ARFs.

\section{Spectral Analysis with Chandra and Suzaku Data}
\label{sec:chandra_ana}

\subsection{Definition of regions for Spectral Analysis}
\label{sub:chandra_region}

The left and right hand panels of figure \ref{fig_ima} show the X-ray source images from Chandra and Suzaku 
overlaid with optical contours constructed from Digitized Sky Survey data. 
Because the starburst region rises about 1.3~kpc ($\sim$15$''$) from the center \citep{cecil_02}, 
we extracted spectra from the following regions of Chandra and Suzaku images. 
For Chandra, there are three regions centered on the nucleus at coordinates 
(R.A., Dec.)=(\timeform{10h01m58.51s}, \timeform{+55D40'49.4''})\@. 
1) a circle with a 0.5$'$ radius (2.25~kpc; hereafter `0.5$'$ circle'); 
2) an inner ring between 0.5$'$--1$'$ (2.25--4.5~kpc; `0.5$'$--1$'$ ring'), and 
3) an outer ring between 1$'$--2$'$ (4.5--9~kpc; `1$'$--2$'$ ring').
The background region for the Chandra analysis is a ring with a radius of between 2$'$--3$'$ 
as shown in figure \ref{fig_ima} (left). 
The count rate in this background ring is less than 5\% of that measured within the central 2$'$.

For Suzaku, the source region is a circle with a 4$'$ radius (`4$'$ circle') 
centered on coordinates (R.A., Dec.)=(\timeform{10h01m59.69s}, \timeform{+55D39'59.8''})\@. 
This is much larger than for Chandra owing to Suzaku's half-power diameter of $\sim$2$'$. 
The background region is the entire XIS field of view, excluding the source region 
and two additional circular apertures with radii of 2$'$ each, 
centered on background field objects as shown in figure \ref{fig_ima} (right). 
Spectral fitting was performed with XSPEC 12.6 \citep{arnaud_96}.

\subsection{Point Sources with Chandra}
\label{sub:point_sources}

The low angular resolution of Suzaku, $\sim$2$'$, prevents separation 
of point sources (PSs) but Chandra is ideal for this purpose. 
The PSs were detected by the ``wavdetect'' CIAO tool task, 
with parameter `sigthresh' set to its default value of 10$^{-6}$ in the 2--10~keV energy range. 
We accumulated the spectra of all detected PSs within 4$'$ of the nucleus 
using regions determined by ``wavdetect'', and fitted their sum with the following model: 
phabs$_{\rm PS}$ $\times$ (power-law$_{\rm PS}$ + Gaussian). 
Here, phabs$_{\rm PS}$ represents photoelectric absorption.
The integrated spectra from PSs are usually approximated by a power-law model, 
denoted above by power-law$_{\rm PS}$ (e.g., \cite{xu_05}).
Since \citet{iyomoto_01} detected a strong Fe-K line at 6.4~keV, 
we also include a Gaussian above. 
This model reproduces the accumulated spectrum well with ${\chi^2}$/d.o.f.=46/36.
The derived parameters are listed in table \ref{tab_point} and the integrated spectrum is shown 
in figure \ref{fig_point}. 
The index of power-law$_{\rm PS}$ is 1.92$_{-0.28}^{+0.15}$, a typical value for AGN spectra.
The central line energy of Gaussian model is 6.30$_{-0.03}^{+0.41}$~keV, which is consistent with that 
of \citet{iyomoto_01} and \citet{fukazawa_11}.

\begin{table}
\caption{
Best-fit parameters for the background regions of Suzaku with  apec components + PL model.$^{\ast}$}
\label{tab_bgd}
\begin{center}
\begin{tabular}{lccc} \hline\hline
Parameters               &             &                          \\ \hline
$N_{\rm H}$               & (10$^{20}$ cm$^{-2}$) & 0.795 (fix)   \\
$\Gamma_{\rm CXB}$        &             & 1.46$_{-0.07}^{+0.08}$            \\ [1.0ex]
$kT_{\rm MWH}$            & (keV)       & 0.30$\pm{0.03}$        \\ 
Abundance                & (solar)     & 1 (fix)                    \\ 
$kT_{\rm LHB}$             & (keV)       & 0.10$\pm{0.01}$         \\ 
Abundance                 & (solar)     & 1 (fix)                   \\ 
$Norm_{\rm LHB}/Norm_{\rm MWH}$  &             & 12.25$_{-4.85}^{+8.30}$        \\[2.0ex] 
${\chi^2}$/d.o.f.         &             & 333/303                   \\ \hline \hline
\end{tabular}
\end{center}
\parbox{\textwidth}{\footnotesize
\footnotemark[$\ast$]
The apec components for spectra in the background region of \\
NGC~3079 with absorbed MWH and LHB components for \\the Galactic emission, 
and a PL model for the CXB.}
\end{table}

\subsection{Estimation of Background Spectra with Suzaku}
\label{sub:suzaku_back}

For the Suzaku data analysis, we performed simultaneous fitting of the source and 
background spectra, instead of subtracting the background as with Chandra. 
The reason is because of the extended tail of point spread function and the larger vignetting 
in the case of Suazaku, as compared to Chandra \citep{serlemitsos_07}. 
Therefore, we first fitted the spectra extracted from the Suzaku background region. 
Since the NXB is already subtracted, the background components consist of 
Cosmic X-ray Background (CXB) and Galactic thermal emission. 
We assumed a power-law model for the CXB component, and a two-temperature thermal plasma model  
for Galactic emission. 
One of the plasmas is the Milky Way Halo (MWH) and the other is the sum of 
the solar wind charge exchange (SWCX) emission and the Local Hot Bubble (LHB) \citep{yoshino_09}.

The fitting was performed with the following model;
phabs$_{\rm G}$ $\times$ (power-law$_{\rm CXB}$ + apec$_{\rm MWH}$) + apec$_{\rm LHB}$.
phabs$_{\rm G}$ represents photoelectric absorption with column density fixed to the Galactic value 
0.795$\times10^{20}$ cm$^{-2}$ in the direction of NGC~3079.
The apec$_{\rm MWH}$ and apec$_{\rm LHB}$ components represent thin thermal plasmas using 
the APEC model \citep{smith_01} from MWH and LHB (and SWCX), respectively. 
The redshift and metal abundance are fixed at zero and one solar, respectively. 
We simultaneously fitted the spectra from BI and FI CCDs in 
the 0.4--5.0 and 0.5--5.0 keV range, respectively.
The derived parameters are listed in table \ref{tab_bgd}.
This model reproduced the spectra approximately, with ${\chi^2}$/d.o.f.=333/303.
Because the photon index and flux of power-law$_{\rm CXB}$ and temperatures of apec models are 
consistent with typical values (e.g. \cite{kushino_02}; \cite{mccammon_02}; \cite{yoshino_09}), 
these models represent the background spectra well. 
We adopt these parameters for our background model.

\renewcommand{\arraystretch}{0.7} %
\begin{table*}
\caption{
Best-fit parameters for the 0.5$'$ circle, 0.5$'$-1$'$ ring, 1$'$-2$'$ ring, and 1$'$ circle regions of Chandra and 4$'$ circle of Suzaku with vapec components, PL model, and Gaussian model.}
\label{tab_sim}
\begin{center}
\begin{tabular}{lccccc} \hline\hline
Parameters & & Chandra  & Suzaku & Chandra  & Suzaku \\ 
           & &           &       & 1$'$ abundance linked  &  \\ \hline
             & & 0.5$'$ circle & & 0.5$'$ circle & \\ 
$N_{\rm H_{G}}$ & (10$^{20}$ cm$^{-2}$) &   0.795 (fix) & 0.795 (fix)&   0.795 (fix) & 0.795 (fix) \\
kT & (keV) &  0.65$^{+0.05}_{-0.04}$ & linked to left&  0.63$\pm0.04$ & linked to left\\
O & (solar) &  0.61$^{+0.41}_{-0.27}$ & linked to left&  0.50$^{+0.24}_{-0.19}$ & linked to left\\
Ne & (solar) &  1.00$^{+0.46}_{-0.34}$ & linked to left&  0.99$^{+0.27}_{-0.23}$ & linked to left\\
Mg, Al & (solar) & 0.46$^{+0.28}_{-0.22}$ & linked to left& 0.48$^{+0.20}_{-0.17}$ & linked to left\\
Si, S, Ar, Ca & (solar) & 0.37$^{+0.29}_{-0.22}$ & linked to left& 0.42$^{+0.24}_{-0.20}$ & linked to left\\
Fe, Ni & (solar) &  0.24$^{+0.08}_{-0.05}$ &  linked to left&  0.22$^{+0.05}_{-0.04}$ &  linked to left\\
Norm$^{\ast}$ & (10$^{-4}$) & 1.90$^{+0.38}_{-0.39}$ &   linked to left& 1.92$^{+0.34}_{-0.31}$ &   linked to left\\
O/Fe & (solar) &  2.56$^{+1.79}_{-1.36}$ &  linked to left&  2.13$^{+1.03}_{-0.77}$ &  linked to left\\
Ne/Fe & (solar) &  4.40$^{+1.85}_{-2.40}$ &  linked to left&  4.50$^{+1.31}_{-1.00}$ &  linked to left\\
Mg/Fe & (solar) &  2.17$^{+1.53}_{-1.41}$ &   linked to left&  2.17$^{+0.91}_{-0.77}$ &   linked to left\\
Si/Fe & (solar) &  1.64$^{+0.74}_{-0.94}$  &  linked to left& 1.90$^{+1.23}_{-0.94}$  &  linked to left\\
total counts & (count) &  2327  &  &  &  \\[1.0ex]
             & & 0.5$'$-1$'$ ring & & 0.5$'$-1$'$ ring \\ 
kT & (keV) &   0.45$^{+0.07}_{-0.06}$ &  linked to left&   0.48$^{+0.07}_{-0.09}$ &  linked to left\\ 
O & (solar) &  0.41$^{+0.41}_{-0.23}$ &  linked to left&  linked to 0.5$'$ &  linked to left\\
Ne & (solar) &    1.07$^{+0.60}_{-0.37}$ &  linked to left&  linked to 0.5$'$ &  linked to left\\
Mg, Al & (solar) &  0.67$^{+0.65}_{-0.39}$ &  linked to left&  linked to 0.5$'$ &  linked to left \\
Si, S, Ar, Ca & (solar) &  1 (fix) & linked to left& linked to 0.5$'$  &  linked to left\\
Fe, Ni & (solar) & 0.23$^{+0.10}_{-0.07}$ &  linked to left&  linked to 0.5$'$ &  linked to left\\
Norm$^{\ast}$ & (10$^{-4}$) &   1.11$\pm0.32$ & linked to left & 1.14$^{+0.26}_{-0.22}$ &  linked to left\\
O/Fe & (solar) &  1.60$^{+2.40}_{-0.80}$ &  linked to left&  linked to 0.5$'$ &  linked to left\\
Ne/Fe & (solar) & 4.60$^{+2.16}_{-1.00}$  & linked to left&  linked to 0.5$'$ &  linked to left\\
Mg/Fe & (solar) &  3.13$^{+1.87}_{-1.84}$  &  linked to left&  linked to 0.5$'$ &  linked to left\\
total counts & (count) &  3898  &  &  &  \\[1.0ex]
             & & 1$'$-2$'$ ring  & & 1$'$-2$'$ ring  &  \\ 
kT & (keV) &  0.24$^{+0.03}_{-0.02}$ & linked to left&  0.24$\pm0.03$ & linked to left \\ 
O & (solar) &  0.30$^{+0.27}_{-0.12}$ & linked to left&  0.31$^{+0.28}_{-0.12}$ & linked to left \\
Ne & (solar) &   1.57$^{+1.48}_{-0.69}$ & linked to left &   1.70$^{+1.56}_{-0.71}$ & linked to left \\
Mg, Al & (solar) &  2.87$^{+4.56}_{-1.93}$ & linked to left&  3.23$^{+3.34}_{-2.07}$ & linked to left  \\
Si, S, Ar, Ca & (solar) &  1 (fix) & linked to left &  1 (fix) & linked to left \\
Fe, Ni & (solar) & 1.22$^{+1.44}_{-0.59}$ & linked to left  & 1.26$^{+1.04}_{-0.62}$ & linked to left  \\
Norm$^{\ast}$ & (10$^{-4}$) &  1.60$^{+0.63}_{-0.64}$ & linked to left&  1.59$^{+0.62}_{-0.64}$ & linked to left \\
O/Fe & (solar) & 0.24$^{+0.47}_{-0.14}$ & linked to left & 0.24$^{+0.29}_{-0.13}$ & linked to left \\
Ne/Fe & (solar) &  1.30$^{+0.58}_{-0.40}$ & linked to left&  1.36$^{+0.64}_{-0.61}$ & linked to left\\
Mg/Fe & (solar) &  2.35$^{+1.46}_{-1.40}$ & linked to left&  2.42$^{+1.79}_{-1.77}$ & linked to left\\[1.0ex]
$N_{\rm H_{PS}}$ &  (10$^{20}$ cm$^{-2}$) &  - & 115 (fix) &  - & 115 (fix) \\
$\Gamma_{\rm PS}$& &  - & 1.92 (fix)&  - & 1.92 (fix) \\ 
total counts$^{\dagger}$ & (count) &  13291  & 63313 &  &  \\[1.0ex]
${\chi^2}$/d.o.f. &  &  & 885/805 &  &886/809  \\ \hline \hline
\end{tabular}
\end{center}
\parbox{\textwidth}{\footnotesize
\footnotemark[$\ast$]
Normalization of the vapec component divided by the solid angle, $\Omega^{\makebox{\tiny\sc u}}$ of the source image in ``xissimarfgen'',
${\it Norm} = \int n_{\rm e} n_{\rm H} dV \,/\,
(4\pi\, (1+z)^2 D_{\rm A}^{\,2}) \,/\, \Omega^{\makebox{\tiny\sc u}}$
$\times 10^{-14}$ cm$^{-5}$~arcmin$^{-2}$, 
where $n_{\rm e}$ and $n_{\rm H}$ is the electron and proton density, respectively, z is the redshift of the source, and $D_{\rm A}$ is the angular diameter distance to the source.\\
\footnotemark[$\dagger$]
Total counts are not subtracted background emission and  is accumulated of all XIS detectors in Suzaku.
}
\end{table*}
\renewcommand{\arraystretch}{1} %

\begin{figure*}
\begin{center}
\centerline{
\FigureFile(\textwidth,\textwidth){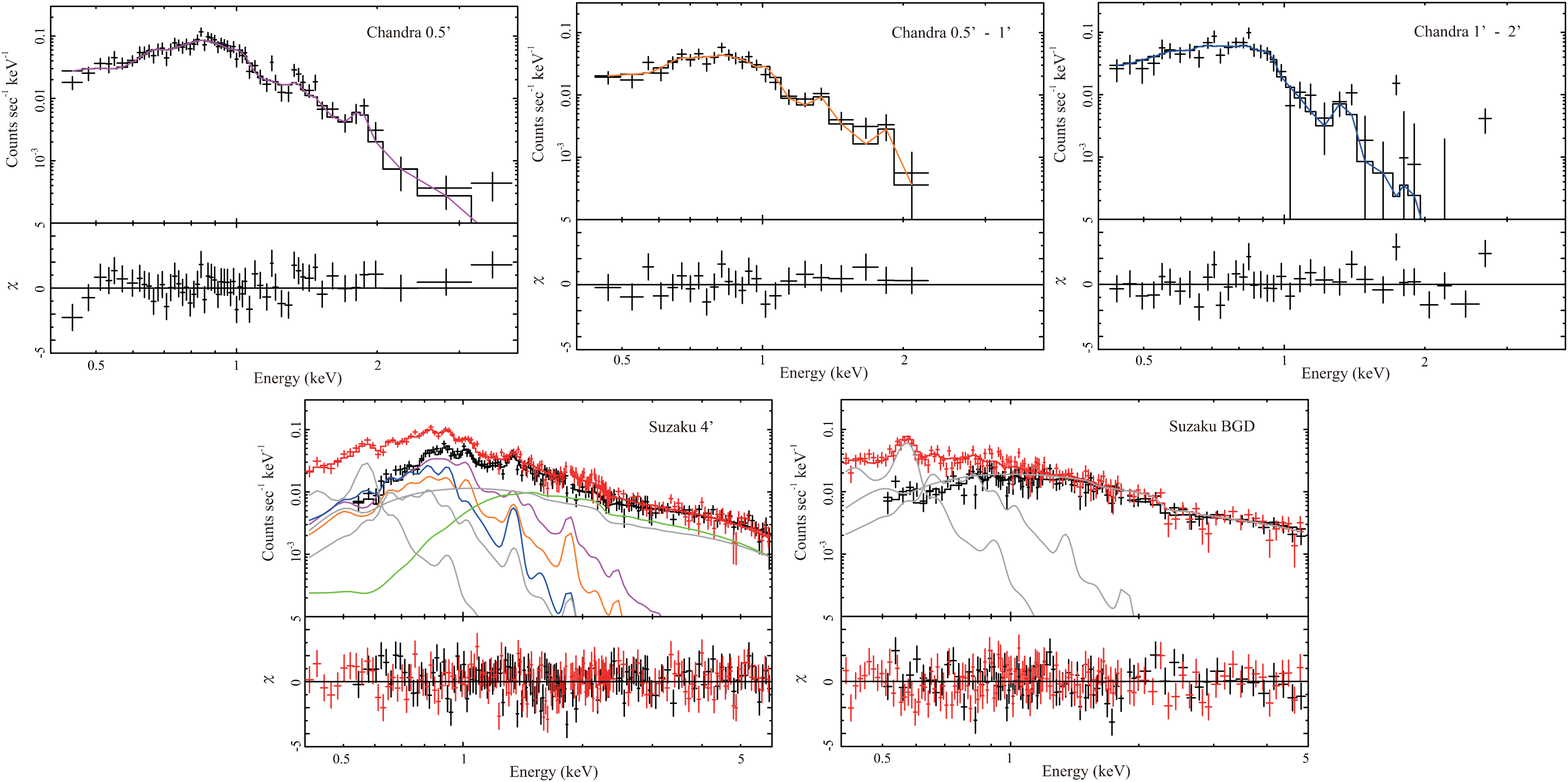}}
\caption{
Simultaneous fitting of Chandra and Suzaku data. 
Top panels: Background-subtracted ACIS-3 spectra of the 0.5$'$ circle (left), 
0.5$'$-1$'$ ring (middle), and 1$'$-2$'$ ring (right) region, 
shown without removal of instrumental response. 
Magenta, orange, and blue lines denote the ISM component (vapec) each region. 
Bottom panels: 
NXB-subtracted XIS0 (black) and XIS1 (red) spectra 
of the 4$'$ circle of NGC~3079 (left panel), and those of the background region (right panel), 
shown without removal of instrumental response. 
Black and red lines show the best-fit model for XIS0 and XIS1, 
respectively. For simplicity, only the model components for XIS1 spectra 
are shown. 
Green line shows accumulated emission from point sources (absorbed power-law$_{\rm PS}$). 
The magenta, orange, and blue lines are linked to Chandra spectra.
The gray lines are the Galactic background emission (apec$_{\rm MWH}$ and apec$_{\rm LHB}$) 
and the CXB components, respectively. 
The background components are common between 
the on-source and background spectra, but scaled to the respective 
data accumulation area.
}\label{fig_sim}
\end{center}
\end{figure*}

\begin{figure*}
\begin{center}
\centerline{
\FigureFile(\textwidth,\textwidth){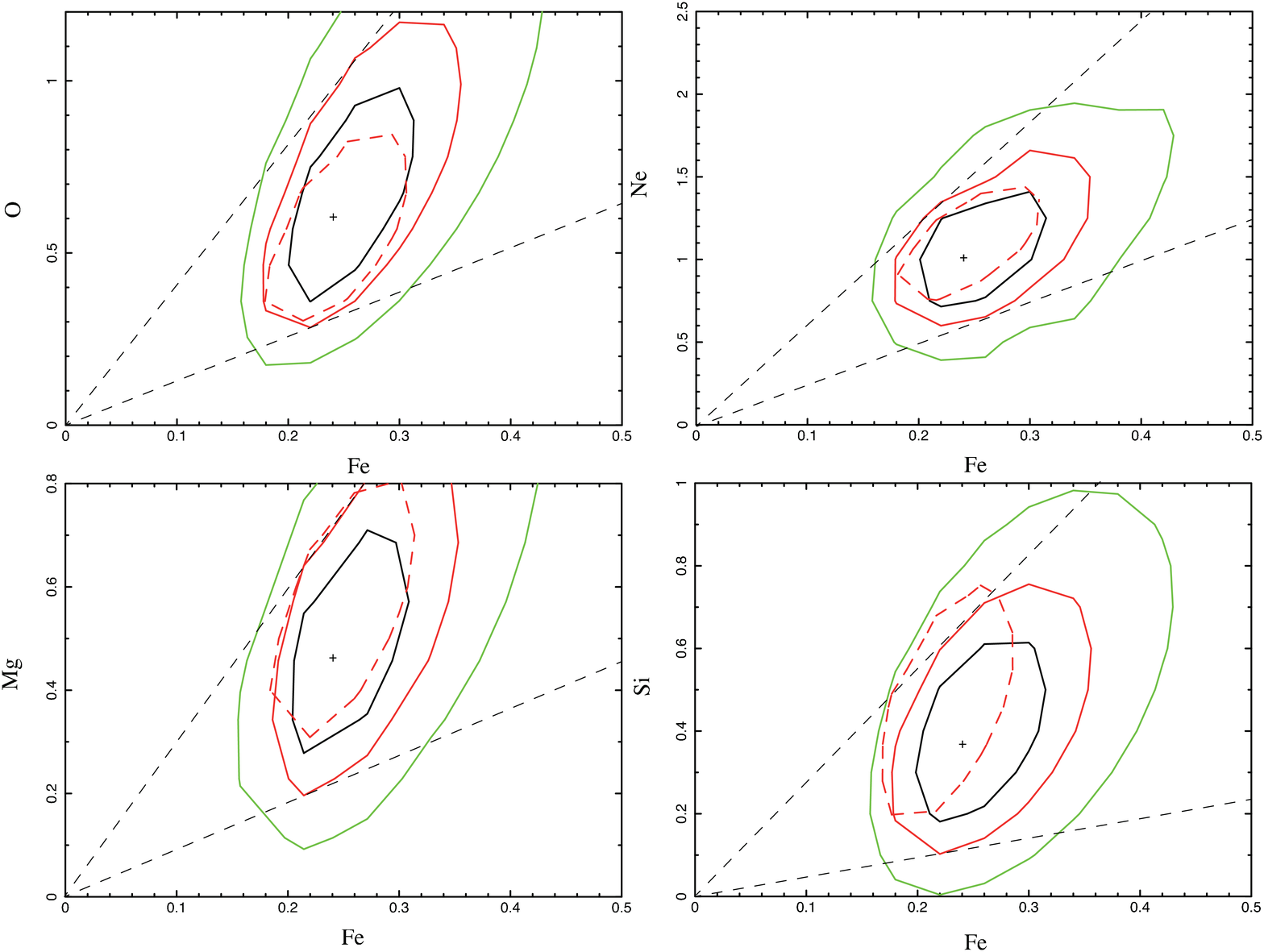}}
\caption{
Confidence contours between metal (O, Ne, Mg, and Si) and Fe 
abundances, determined for 0.5$'$ circle region. 
The black, red, and green contours represent 68\%, 90\%, and 99\% confidence ranges, 
respectively. 
The black dashed lines indicate 90\%-confidence abundances of these metals relative to Fe.
The red dashed contours show the 90\% confidence ranges for when the abundances in the 0.5$'$ 
circle and 0.5$'$--1$'$ ring regions are linked to each other. 
}\label{fig_cont}
\end{center}
\end{figure*}

\subsection{Simultaneous Fit of Chandra and Suzaku Data}
\label{sub:suzaku_fit}

We fitted simultaneously the spectra of the three regions 
(0.5$'$ circle, 0.5$'$--1$'$ ring, and 1$'$--2$'$ ring) of Chandra background-subtracted data, 
along with the spectra from the 4$'$ circle and background region of Suzaku. 
Each Chandra spectrum was fitted with 
phabs$_{\rm G}$ $\times$ vapec$_{\rm 0.5\timeform{'}~or~0.5\timeform{'}-1\timeform{'}~or~1\timeform{'}-2\timeform{'}}$. 
The phabs$_{\rm G}$ factor is fixed to the Galactic value 0.795$\times10^{20}$ cm$^{-2}$.
The vapec$_{\rm 0.5\timeform{'}~or~0.5\timeform{'}-1\timeform{'}~or~1\timeform{'}-2\timeform{'}}$ model represents 
thermal plasma emission in the ISM of NGC~3079, 
corresponding to the 0.5$'$ circle, 0.5$'$--1$'$ ring, and 1$'$--2$'$ ring regions of Chandra, respectively. 
We divided the metals into the groups of O, Ne, (Mg \& Al), (Si, S, Ar, \& Ca) and 
(Fe \& Ni) based on the metal synthesis mechanism of SN, allowing each group to vary. 
However, the value of the Si group is fixed to 1 solar beyond the 1$'$ circle region.

For the Suzaku 4$'$ circle spectra, we included all three Chandra components in the models.
We fitted the spectra of Suzaku 4$'$ circle with following model: 
phabs$_{\rm G}$ $\times$ (vapec$_{\rm 0.5\timeform{'}}$ + vapec$_{\rm 0.5\timeform{'}-1\timeform{'}}$ + 
vapec$_{\rm1\timeform{'}-2\timeform{'}}$) 
+ phabs$_{\rm PS}$ $\times$ (power-law$_{\rm PS}$ + Gaussian) + background models.
The first term represents the three components of Chandra regions, which were linked 
to that of corresponding each Chandra region. 
The accumulated emission from PSs including the nucleus is modeled as 
phabs$_{\rm PS}$ (power-law$_{\rm PS}$ + Gaussian). 
The phabs$_{\rm PS}$ and index of power-law$_{\rm PS}$ are fixed to the best fit values 
derived in subsection \ref{sub:point_sources}. 
The parameters of the background models were linked between the 4$'$ circle and Suzaku background regions. 
An energy range of 0.4--4.0~keV is used for the 0.5$'$ circle, 0.5$'$--1$'$ ring, and 1$'$--2$'$ ring 
of Chandra, and from 0.4 (BI) or 0.5 (FI) to 5.0~keV and 6.0~keV for the background and 4$'$ circle of Suzaku, 
respectively.
Because there are known calibration issues of the Si edges in all XIS CCDs, 
the energy range of 1.84--1.86~keV is ignored \citep{koyama_07} in Suzaku spectra.

The fitted spectra are shown in figure \ref{fig_sim} and the derived parameters are listed in 
table \ref{tab_sim}.
This model reproduced these spectra approximately, with ${\chi^2}$/d.o.f.=885/805. 
The derived temperatures of thermal models are 0.65, 0.45, and 0.24~keV in 0.5$'$ circle, 
0.5$'$--1$'$ ring, and 1$'$--2$'$ ring regions, respectively. 
Several emission lines seen around 0.5--0.6 keV, 0.6--0.7 keV, 
$\sim$1.3 keV, and $\sim$1.8 keV are identified with K$\alpha$ lines of 
O \emissiontype{VII}, O \emissiontype{VIII}, 
Mg \emissiontype{XI}, and Si \emissiontype{XIII}, respectively. 
The emission bump around 0.7--1 keV corresponds to 
the Fe-L complex, as well as to K-lines from Ne \emissiontype{IX} and 
Ne \emissiontype{X}. 
The metal abundances in all three regions have very similar best-fit values, though with large uncertainties. 
In particular, the lowest temperature (0.24~keV) component makes the faintest 
contribution to the overall spectrum, so the low O abundance in this region should have a 
large associated systematic error. 
Meanwhile, the O abundance derived from highest temperature (0.65~keV) 
bright component is more reliable.

In order to examine the abundance ratios rather than their absolute vales, 
we calculated the confidence contours between the abundance of various metals 
(O, Ne, Mg, and Si) to that of Fe, in order to derive abundance ratios. 
The results are shown in table \ref{tab_sim} and figure \ref{fig_cont}, 
where we also indicate 90\%-confidence abundance of these metals 
relative to Fe with black dashed line. 
The elongated shape of the confidence contours indicates that the relative values 
were determined more accurately than the absolute values. 
Because the abundance patterns between the 0.5$'$ circle and 0.5$'$-1$'$ ring 
are consistent with each other, we fitted the Chandra and Suzaku spectra again, 
this time setting the abundances to be common between these two regions. 
The derived parameters agree with the results in previous free fit, as summarized 
in table \ref{tab_sim} and figure \ref{fig_cont}, but now the abundances in the inner regions are 
better constrained. 
In figure \ref{fig_cont}, we only plotted the 90\%-confidence abundance of these metals 
relative to Fe with red dashed line.
The calculated abundance ratios of O/Fe, Ne/Fe, Mg/Fe, and Si/Fe, 
are 2.13$^{+1.03}_{-0.77}$, 4.50$^{+1.31}_{-1.00}$, 2.17$^{+0.91}_{-0.77}$, and 1.90$^{+1.23}_{-0.94}$ 
within the 1$'$ circle, respectively.

\subsection{Systematic Uncertainties in the Abundance Ratios}
\label{sub:systematic}

In Chandra analysis, we extracted the spectra after excising the PSs. 
We fitted the spectra by setting the power-law model normalization within 
the 0.5' circle region to be free, in order to investigate the effect of 
the emission from PSs below the detection threshold.
The flux upper limit of the power-law model corresponds to $<$20\% 
of the total flux in the central region.
Furthermore, the parameters of vapec model do not change significantly.

In order to investigate systematic uncertainties for the abundance ratios, 
we set free the absorption parameter, phabs$_{\rm G}$, because additional absorption may 
arise along the edge-on line-of-sight within the ISM of NGC~3079.  
We tried setting free the absorption affecting only the central 0.5$'$ circle region, 
where intrinsic absorption should be the highest. 
The resultant absorption was constrained to an upper limit of 8.4$\times10^{20}$ cm$^{-2}$. 
None of the other parameters changed significantly, implying that the systematic effect 
of the absorption uncertainty is almost negligible.

Next, we investigated abundance trends amongst the $\alpha$-elements. 
In particular, we set the abundances of Ne and Mg to be equal to that of O in each region 
because we expect similar distributions among these. 
In the 0.5$'$ circle and 0.5$'$-1$'$ ring regions, the ratios of O/Fe, Ne/Fe, and 
Mg/Fe are consistent to those of each region, as in subsection \ref{sub:suzaku_fit} above. 
On the other hand, in the 1$'$-2$'$ ring region, the ratio of O/Fe increased to near solar. 
As mentioned in subsection \ref{sub:suzaku_fit}, 
the O abundance of the lowest temperature component may have a large 
systematic error owing to its faintness.

The Fe-L lines have systematic uncertainties related to unknown atomic physics, 
as discussed in \citet{matsushita_00}. 
When convolving with the XIS response matrix, differences in the Fe-L lines between APEC \citep{smith_01} 
and MEKAL \citep{mewe_85} models are typically of order 10-20\% \citep{matsushita_00}. 
Therefore, we re-fitted the spectra in the same way as above, but with an inclusion of 15\% 
systematic errors over the energy range of 0.75--1.2 keV. 
The ${\chi^2}$/d.o.f. improves to 603/623, with null-hypothesis probability of $\sim$0.7. 
The derived parameters in this case are completely consistent to those of 
subsection \ref{sub:suzaku_fit}.
Therefore, these results are useful to assess whether the temperatures 
and metal abundances in the ISM were reasonably determined or not.
Furthermore, the APEC code has recently been updated to version 2. The ratios among L-lines complex of Fe 
have changed significantly from version 1. 
We have fitted the spectra with version 2 of APEC code, and found that all resultant 
parameters are consistent with those of table \ref{tab_sim}, with ${\chi^2}$/d.o.f.=893/805. 
In this statistic, there are no differences between results in version 1 and 2 of APEC code.

\begin{figure*}
\begin{center}
\centerline{
\FigureFile(\textwidth,\textwidth){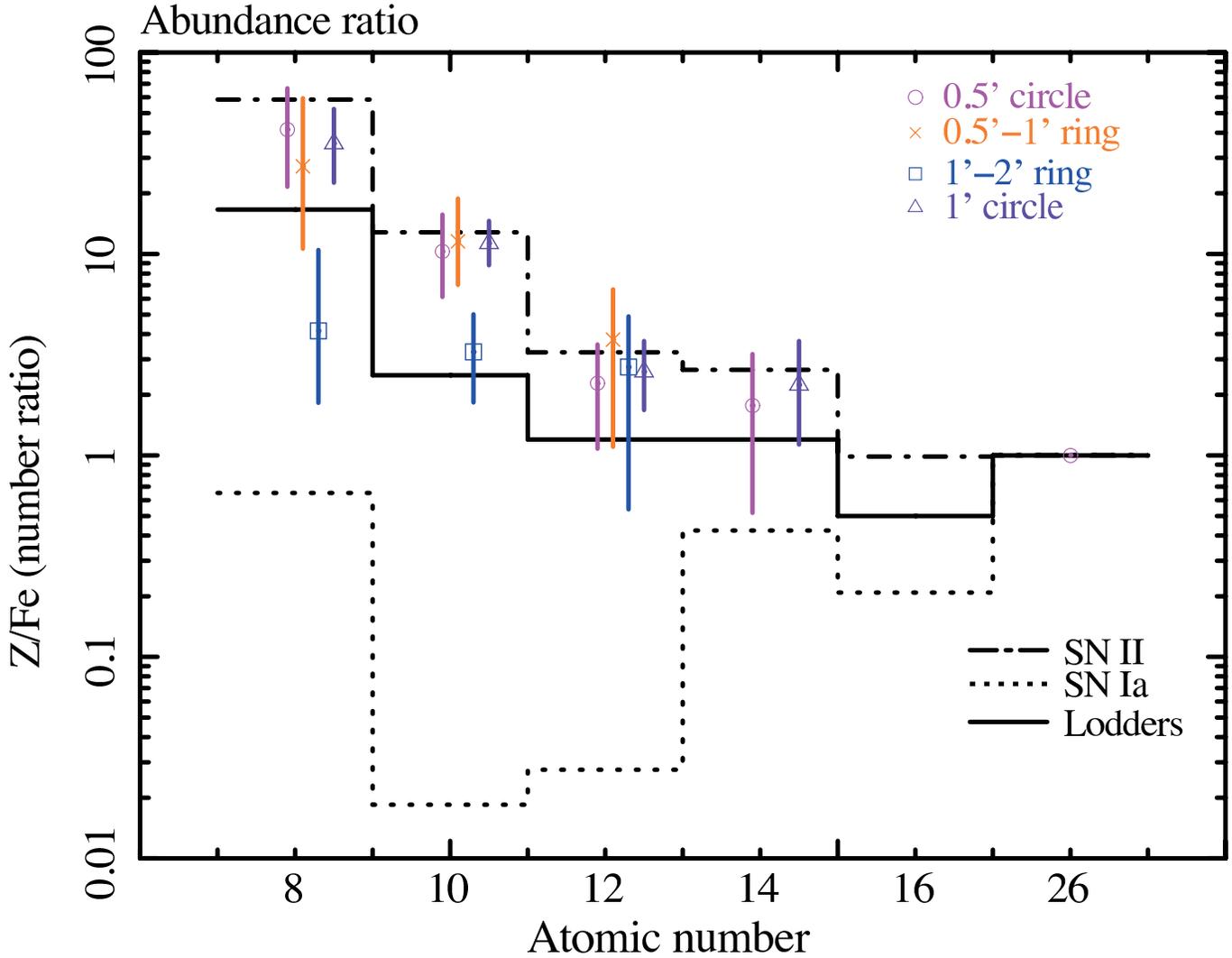}}
\caption{
Abundance ratios of O, Ne, Mg, and Si to Fe for the best-fit model of NGC~3079 
derived by simultaneously fitting the Chandra and Suzaku spectra. 
Magenta, orange, blue, and purple data show the abundance patterns of 0.5$'$, 0.5$'$--1$'$, 1$'$--2$'$, and 1$'$, 
respectively. 
Solid, dot-dashed, and dotted lines represent the number ratios of metals to 
Fe for solar abundance, for SN II, and for SN Ia products \citep{lodders_03, iwamoto_99, nomoto_06}, respectively. 
}\label{fig_num}
\end{center}
\end{figure*}

\begin{table*}
\caption{
Spatial scales of well-known starburst galaxies, in which abundance patterns are SN II--like. The distance scale is measured perpendicular to the disk in each case.}
\label{tab_wind}
\begin{center}
\begin{tabular}{lccccc} \hline\hline
Galaxy & & NGC~3079 & NGC~253$^{\ast}$ & NGC~4631$^{\dagger}$  & M~82 \\ \hline
D & (kpc) &  4.5 & 9 & 13 & 11$^{\ddagger}$ \\ \hline \hline
\end{tabular}
\end{center}
\parbox{\textwidth}{\footnotesize
\footnotemark[$\ast$]
\citet{bauer_08}
\footnotemark[$\dagger$]
\citet{yamasaki_09}
\footnotemark[$\ddagger$]
\citet{tsuru_07}
}
\end{table*}

\section{Discussion}\label{sec:discuss}

\subsection{Spatial Metal Abundance Patterns}

We have successfully derived the spatial metal abundance patterns 
in the hot ISM of NGC~3079, 
using both Chandra and Suzaku for the first time. 
The combination of these missions enables us to optimally utilize the advantages of 
both good spatial and good spectral resolution at high sensitivity. 
Figure \ref{fig_num} shows the abundance patterns in the 0.5$'$ circle, 0.5$'$--1$'$ ring, 1$'$--2$'$ ring, 
and 1$'$ circle regions. 
The abundance patterns of NGC~3079 can now be compared with the solar abundance table \citep{lodders_03} 
and those expected from SN II and SN Ia yields.  
The SN II yields computed by \citet{nomoto_06} are used here. These refer to an average over 
the Salpeter initial mass function for stellar masses from 10 to 50 $M_{\odot}$, with a progenitor 
metallicity of $Z=0.02$\@.
The SN Ia yields from the W7 model of \citet{iwamoto_99} were adopted.
Within a 4.5~kpc region (i.e. a circle with radius of 1$'$),  
the abundance pattern is consistent to that of SN II within the uncertainties. 
We thus infer this region to be significantly enriched by SN II metal yields, undoubtedly related to 
the starburst activity in this system. 
On the other hand, the abundance pattern is not SN II--like beyond 4.5~kpc, 
presumably because the starburst has had little impact here. 
This emission comes from the intrinsic galactic halo and is not enriched by the starburst 
outflow yet.

Using AKARI, \citet{yamagishi_10} detected extended emission from hot dust ($\sim$30~K) 
in the center regions (4~kpc) of NGC~3079, possibly associated with the starburst activity. 
The emission ratios of ionized to neutral polycyclic aromatic hydrocarbons are 
higher in the center ($\sim$2~kpc) than in the disk region. 
Both facts indicated that star formation is relatively active in the center region, 
consistent with our results based on X-ray data.

\subsection{Time Scale of Starburst Activity}
\label{subsec_time}

The total infrared luminosity of NGC~3079 indicates a rather low star-formation rate, and measured 
gas-to-dust mass ratios over the galaxy are unusually high, which led \citet{yamagishi_10} to suggest 
that NGC~3079 is still in a rather early phase of the starburst, 
with nuclear star formation activity expected to increase in the near future. 
Derived abundance patterns indicate that the ISM within at least 4.5~kpc from the center 
contains metals synthesized by SN II. Using this fact, the approximate time ($t$)
elapsed since starburst activity began can be estimated by computing the distance ($R$) 
travelled by the wind according to the numerical hydrodynamic starburst wind simulations of 
\citet{tomisaka_93}. $R$ can be written as\\
\begin{eqnarray*}
R(t)~\simeq 31~\rm{kpc}~(\frac{n_{halo}}{2 \times 10^{-2}~\rm{cm^{-2}}})^{-1/5}~\times~\\
(\frac{L}{3 \times 10^{42}~\rm{erg~s^{-1}}})^{1/5}(\frac{t}{50~\rm{Myr}})^{3/5}
\end{eqnarray*}
where $L$ is the luminosity and $n_{\rm halo}$ is the particle density of hot plasma \citep{tomisaka_93}.
From the Suzaku spectra, we measure a 0.3--5 keV total unabsorbed ISM luminosity of 
7.33$\times$10$^{39}$ erg s$^{-1}$. 
The plasma (electron) density is derived from the normalization of vapec model, 
${\it Norm} = \int n_{\rm e} n_{\rm H} dV \,/\,
(4\pi\, (1+z)^2 D_{\rm A}^{\,2})$
$\times 10^{-14}$ cm$^{-5}$~arcmin$^{-2}$, 
where $D_{\rm A}$ is the angular distance to the source.
The normalization of vapec model within 1$'$ circle region is 3.0 $\times$ 10$^{-4}$  
cm$^{-5}$~arcmin$^{-2}$ (table \ref{tab_sim}).
Considering the ratio of proton to He in plasma for SN II--like abundances, 
we use a particle density of 5 $\times$ 10$^{-3}$ cm$^{-3}$, as obtained from the emission measure 
defined by $\int 2.67 n_{\rm e} n_{\rm H} dV$. 
With these assumptions, we derived the time for the wind to travel a distance of 4.5~kpc as 10~Myr. 
In combination with the 4~kpc extent of the infrared--emitting starburst region, 10~Myr may 
be considered as an upper limit for the age of the starburst. 

In two-dimensional numerical hydrodynamical simulations incorporating more realistic spatial 
plasma density distribution (e.g. \cite{tomisaka_93}, \cite{strickland_00}), 
starburst-driven galactic winds can reach $\sim$ 5~kpc after 5 Myr $<$ $t$ $<$ 10 Myr. 
Our calculation is comparatively crude (assuming spherical symmetry and a uniform density), 
but the results are qualitatively consistent.

\subsection{Comparison of the Starburst Phases among Starburst Galaxies}

In a few other well-known starburst galaxies, NGC~4631, M~82, and NGC253, 
the metal abundances of the outflow regions also show $\alpha$-element enhancements relative 
to Fe, consistent with enrichment by SN II
(\cite{yamasaki_09}, \cite{tsuru_07}, \cite{konami_11}, \cite{mitsuishi_12}). 
The central region of NGC~3079 shows a similar pattern. 
In table \ref{tab_wind}, we summarize the spatial extents of the regions with 
SN II--like patterns in these galaxies, measured perpendicular to the disk in each case. 
Spatial scales of order $\sim$10~kpc in NGC~4631, M~82, and NGC253 indicate that 
more than 10~Myr have passed since starburst activity began, according to the simulations of \citet{tomisaka_93}.
By comparison, the smaller spatial extent and age upper-limit inferred in subsection \ref{subsec_time} 
indicate that NGC~3079 is in an earlier phase of activity. 
Though the SN II--like abundance patterns have been observed through starburst galaxies, 
the different spatial distributions of those patterns correspond to their starburst activity phases.

Though the temperature of dust in the center of NGC~3079 is about $\sim$ 30~K, 
in the central ($<$1~kpc) region of M~82 at least two components with higher dust 
temperatures of $\sim$45~K and 160~K have been found 
(\cite{yamagishi_10}, \cite{gandhi_11}). 
These may result from prolonged heating associated with the longer period of starburst 
activity in M~82, as compared to NGC~3079. 
Copious production of dust in more mature starbursts like M~82 also ought to 
absorb a larger fraction of the starburst radiative output and thus 
become more heated. NGC~3079 instead displays a comparatively low dust-to-mass ratio at present. 
For NGC~253, \citet{engelbracht_98} concluded that the source is in a late starburst phase, 
based upon detection of extended infrared emission. 
Modeling of this emission implied an age of 20 to 30 Myr. 
In NGC~4631, the ratios of $^{12}$CO(2-1)/$^{12}$CO(1-0) are similar to the average value seen in the 
nuclei of normal galaxies \citep{golla_94}. 
The authors used this fact to suggest that source was undergoing mild central star formation. 
But these ratios could also be consistent with a late phase of starburst activity.

The abundance pattern beyond 4.5~kpc in NGC~3079 is rather consistent with solar \citep{lodders_03}. 
The disk region in the starburst galaxy NGC~4631 similarly shows solar abundance. 
It is interesting that the hot ISM in NGC~4258, a spiral galaxy without starburst activity \citep{konami_09}, 
also behaves similarly. 
These facts suggest that the metallicity of the ISM both before, and after an era of intense starburst activity, 
may look quite similar.

\section{Conclusion}\label{sec:conclusion}

We have performed X-ray spectral analysis of NGC~3079 with Chandra and Suzaku. 
The temperatures in the ISM thermal plasma are found to show a negative radial gradient. 
In the central ($<$4.5~kpc) region, the metal abundance patters are 
consistent with those synthesized by SN II. 
On the other hand, abundances in the outer region ($>$4.5~kpc) are closer to solar. 
These indicate that the metal abundances in the center have been enriched 
with SN II yields, but that the outer region has not yet been influenced by the starburst.
This is consistent with starburst activity in NGC~3079 being in a relative-early evolutionary phase.

\bigskip
We thank the referee for providing valuable comments.
We gratefully acknowledge all members of the Suzaku hardware and software 
teams and the Science Working Group. 
SK is supported by JSPS Research Fellowship for Young Scientists.

\end{document}